\documentclass[acmsmall,screen,nonacm]{acmart}
\AtBeginDocument{%
  }

\setcopyright{acmlicensed}
\copyrightyear{2018}
\acmYear{2018}
\acmDOI{XXXXXXX.XXXXXXX}
\acmConference[Conference acronym 'XX]{Make sure to enter the correct
  conference title from your rights confirmation email}{June 03--05,
  2018}{Woodstock, NY}
\acmISBN{978-1-4503-XXXX-X/2018/06}

\usepackage{tabularx}
\usepackage{multirow}
\usepackage{array}
\usepackage{bm}
\usepackage{enumitem}
\setlist[enumerate,1]{left=0pt}
\setlist[itemize,1]{left=0pt}
\usepackage{cleveref}
\crefname{section}{§}{§§}
\Crefname{section}{§}{§§}
\usepackage{caption}
\usepackage[linesnumbered,ruled,vlined]{algorithm2e}
\SetAlFnt{\small}
\SetAlCapFnt{\small}

\begin{document}

\title{Password Strength Detection via Machine Learning: Analysis, Modeling, and Evaluation}

\author{Jiazhi Mo}
\authornote{Both authors contributed equally to this research.}
\author{Hailu Kuang}
\authornotemark[1]
\email{hailukuang@hainanu.edu.cn}
\affiliation{%
  \institution{Hainan University}
  \city{Haikou}
  \country{China}}

\author{Xiaoqi Li}
\affiliation{%
  \institution{Hainan University}
  \city{Haikou}
  \country{China}}
\email{csxqli@ieee.org}

\renewcommand{\shortauthors}{Mo et al.}

\begin{abstract}
  As network security issues continue gaining prominence, password security has become crucial in safeguarding personal information and network systems. This study first introduces various methods for system password cracking, outlines password defense strategies, and discusses the application of machine learning in the realm of password security. Subsequently, we conduct a detailed public password database analysis, uncovering standard features and patterns among passwords. We extract multiple characteristics of passwords, including length, the number of digits, the number of uppercase and lowercase letters, and the number of special characters. We then experiment with six different machine learning algorithms: support vector machines, logistic regression, neural networks, decision trees, random forests, and stacked models, evaluating each model's performance based on various metrics, including accuracy, recall, and F1 score through model validation and hyperparameter tuning. The evaluation results on the test set indicate that decision trees and stacked models excel in accuracy, recall, and F1 score, making them a practical option for the strong and weak password classification task. 
\end{abstract}




\maketitle

\section{INTRODUCTION}
The rapid advancement of Internet technology has significantly transformed daily life, facilitating access to information and enhancing connectivity like never before. The acceleration of information transmission, the diversification of access channels, and the emergence of big data, blockchain, and the Internet of Things (IoT) have rendered cyberspace an important barrier to national security \cite{vakulyk2020cybersecurity}. Amid these developments, the "Internet Plus" business model has experienced unprecedented growth, increasingly allowing individuals to integrate their lives with digital platforms and assets. Nonetheless, as this integration deepens, network security concerns have become increasingly pronounced, with issues such as data breaches, hacking incidents, and online fraud presenting substantial risks. Notably, in 2011, several websites fell victim to hacking attacks, exposing vast amounts of user data, including the leak of over 6 million user records from the CSDN website \cite{csdnleak2025}.

Among the myriad network security challenges, password security remains a critical area of focus. As a fundamental component of Internet security, the integrity of passwords is intrinsically linked to both personal privacy and organizational security. Cyber attackers utilize various methods, including brute-force cracking, dictionary attacks, and social engineering tactics, to compromise system passwords and unlawfully obtain sensitive user information \cite{castelluccia2012adaptive}. Such attacks pose serious threats to individual privacy and the security of enterprises.

The insecure practices of users selecting passwords have significantly compromised their effectiveness, leading to passwords that often fail to meet theoretical strength standards \cite{adams1999users}. Common issues include the use of overly simplistic passwords and the tendency to reuse passwords across multiple accounts \cite{dell2010password} \cite{das2014tangled}. This practice of password reuse exposes user accounts to "crash attacks," \cite{singer2013rethinking} thereby providing attackers with greater opportunities to exploit vulnerabilities. Moreover, many web application service providers continue implementing outdated password security measures that are ill-equipped to combat evolving attack, resulting in significant data leaks and exacerbating the password security crisis \cite{florencio2010security}. Consequently, it is imperative to investigate strategies for enhancing password security and mitigating various network attacks to protect personal privacy and uphold the security of network systems.

This research aims to analyze data from public password repositories to understand the characteristics and prevalent attack patterns associated with passwords, thereby enhancing the analysis of password security. This study develops a machine learning-based classifier to distinguish between strong and weak passwords, facilitating users' creation of more secure password options. Furthermore, based on the findings of this research, targeted defense strategies are proposed to bolster both user and organizational capacities in password security protection.

The contribution of this paper can be summarized as follows:
\begin{itemize}
\item We examined data from public password databases, elucidating their attributes and standard attack methods to heighten password security awareness and adopt effective password creation.
\item We developed machine learning-based classifiers to evaluate strong and weak passwords, allowing users to assess their passwords during the creation and help them with more robust options.
\item We introduced system password cracking attack and defense strategies and provided defense strategy recommendations to improve password security in network systems.
\end{itemize}

This paper is organized into seven sections. Section \ref{sec::background} introduces the research background, objectives, and significance, followed by a discussion of password-cracking attacks and defensive measures. Subsequently, Section \ref{sec::data_collection} introduces the collection and preprocessing of the public database, and Section \ref{sec::characterization}
analyzed their characterization. Then, a machine learning-based classifier for strong and weak passwords will be introduced in Section \ref{sec::classifier}, including training and experimental evaluation. Finally, Section \ref{sec::defense_strategy} proposes recommendations for password security and defensive strategies, concluding with a summary and future outlook for this research in Section \ref{sec::conclusion}.

\section{PASSWORD CRACKING ATTACK AND DEFENCE}
\label{sec::background}

\subsection{Password Cracking Attack}

\textit{\textbf{Brute-Force Attacks.}}
Brute-force cracking is a technique utilized to compromise a password by exhaustively attempting all possible combinations of characters \cite{najafabadi2014machine}. This method, while theoretically capable of breaking any password, is significantly influenced by the length of the password and the size of the character set employed. With advancements in computing power, particularly the advent of supercomputers, the threat posed by brute-force cracking is escalating, especially for shorter or simpler passwords. Attackers frequently leverage hardware resources such as multi-core processors and Graphics Processing Units (GPUs) to enhance the efficiency of brute-force attacks. Nevertheless, should the password be sufficiently long and complex, brute-force cracking may still require an extensive amount of time to yield results.

\textit{\textbf{Dictionary Attacks.}}
A dictionary attack employs a predetermined list of potential passwords, often referred to as a dictionary, which typically comprises a substantial number of common passwords curated from various sources \cite{bovsnjak2018brute}. The attacker methodically tests each entry in the dictionary until the correct password is identified. In contrast to brute-force cracking, dictionary attacks can be executed with greater speed; however, the efficacy of such attacks is contingent upon the quality of the dictionary and the intricacy of the target password. To enhance the likelihood of success, attackers frequently employ customized dictionaries that encompass common words, phrases across different languages, and potential special characters used by individuals. Additionally, attackers may tap into databases of previously compromised passwords obtained from the internet, utilizing high-frequency passwords and patterns to boost the probability of successful compromise.

\textit{\textbf{Rainbow Table Attacks.}}
Rainbow table attacks utilize pre-calculated hash values stored in a "rainbow table" to facilitate password cracking  \cite{bonneau2012quest}. The primary advantage of this method over brute-force cracking lies in its ability to strike a balance between time efficiency and resource consumption, allowing for the cracking of complex passwords in a relatively expedited manner. Nonetheless, the creation and maintenance of rainbow tables necessitate considerable computational power and significant storage resources. The size of a rainbow table correlates directly with the character set it encompasses and the length of the targeted password; therefore, for particularly intricate passwords, the storage requirements may be substantial. In contemporary contexts, the ongoing enhancement of hashing algorithms and the implementation of salting have begun to mitigate the effectiveness of rainbow table attacks.

\textit{\textbf{Social Engineering.}}
A social engineering attack refers to a technique employed to acquire a user's password by leveraging interpersonal and psychological principles  \cite{weir2010testing}. The attacker manipulates the user into inadvertently disclosing their password through deception and persuasion. This type of attack is often tailored to specific individuals, rendering it more challenging to mitigate. Attackers may gather users' personal information through search engines, social media platforms, and other sources, analyzing this information to devise targeted fraudulent strategies. For instance, an attacker might impersonate a family member, friend, colleague, or superior, create an information void, and solicit a password from the user via telephone, social media, or email. To mitigate the risk of social engineering attacks, users should enhance their security awareness, maintain vigilance regarding messages from unknown or unreliable sources, and necessitate verification of the other party's identity in various forms prior to disclosing passwords.

\textit{\textbf{Hybrid Attacks.}}
Hybrid attacks employ a combination of multiple password-cracking methodologies previously delineated \cite{narayanan2005fast} \cite{shay2016designing}. For example, an attacker may initially attempt to penetrate common passwords utilizing a dictionary attack, subsequently employing brute force on remaining passwords that remain unbroken. This approach capitalizes on the strengths of diverse password-cracking methods, thereby enhancing the overall likelihood of success. Hybrid attacks may also encompass rule-based tactics, whereby the attacker generates password combinations predicated on specific rules, such as appending numbers or special characters to a dictionary list. This strategy effectively addresses the countermeasures users may implement to fortify their passwords, thus increasing the probability of a successful breach.

\textit{\textbf{Crash Attacks.}}
Hackers compile vast arrays of usernames and passwords from data breaches across the internet, subsequently utilizing this information to gain access to other web services. A considerable number of users tend to employ identical usernames and passwords across multiple platforms, resulting in a tangible risk that hackers may successfully log into additional sites or systems employing the compromised credentials. When these systems are associated with financial accounts, there exists a heightened risk of financial theft. Furthermore, some hackers engage in the sale of accounts and passwords that grant unauthorized access to various online platforms.

\textit{\textbf{Phishing Emails.}}
Phishing emails constitute a form of cyber fraud wherein attackers deceive victims by disseminating emails that masquerade as legitimate communications from trustworthy organizations, such as banks, social networking sites, or e-commerce platforms. These emails typically purport to address specific issues, aim to extract personal information, instill fear or panic, and entice victims into clicking on malicious links, divulging sensitive information—such as usernames, passwords, or credit card details—or downloading harmful attachments.

\textit{\textbf{Network Traffic Monitoring.}}
Network traffic monitoring involves the practice of capturing and analyzing packets of network data to extract information regarding network interactions. Monitoring can reveal instances where passwords are transmitted in plaintext over the network, posing significant risks as attackers can intercept and obtain a user's credentials. By scrutinizing passwords within network traffic, it is feasible to identify user accounts relying on weak passwords. Furthermore, monitoring may expose instances of password reuse across multiple communications, increasing vulnerability; when one credential is compromised, hackers may leverage it to gain unauthorized access to other sites or systems.

\subsection{Password Defense}

\textit{\textbf{Password Complexity Requirements.}}
The implementation of password complexity requirements constitutes a critical strategy for ensuring that user-generated passwords possess sufficient strength to withstand various cracking attacks \cite{sasse2001transforming}. Such complexity requirements typically encompass factors such as the overall length of the password, the inclusion of diverse character types (including upper and lower case letters, numerical digits, and special characters), and restrictions on commonly used words. By enforcing these requirements, organizations can significantly reduce the likelihood of successful password breaches through methods such as brute force or dictionary attacks. Businesses and organizations should devise and enforce comprehensive password policies that promote the creation of secure passwords by employees.

\textit{\textbf{Periodic Password Changes.}}
Implementing periodic password changes represents an effective defense strategy aimed at mitigating risks associated with prolonged use of the same password. When users are encouraged to update their passwords regularly, the potential window of opportunity for attackers to access protected resources is minimized, even in instances where an old password is compromised \cite{hao2022sok}. Organizations should establish clearly defined password change intervals that mandate employees to refresh their passwords at specific times \cite{bellovin1992encrypted}. However, these intervals should not be excessively brief, as this may lead to the selection of simplistic passwords or instances of forgotten credentials.

\textit{\textbf{Limiting the Number of Login Attempts.}}
Establishing limits on the number of login attempts is a proven strategy for countering brute-force and dictionary attacks. In cases where a user fails to log in after several attempts successfully, the system may temporarily lock the account or extend the waiting period between attempts, thus preventing unauthorized individuals from making continuous attempts to access the account. Furthermore, the system can send notifications to users, alerting them to potential attempts at unauthorized access \cite{pinkas2002securing}. It is important to recognize that while limiting login attempts enhances security, it may also inadvertently hinder legitimate users from accessing their accounts due to misremembered passwords. Therefore, a careful balance between security measures and user experience should be maintained.

\textit{\textbf{Password Hashing and Salting.}}
In the context of password storage, employing encryption through a hashing function is a widely accepted security practice. This function transforms the password into a fixed-length hash value, rendering it infeasible for an attacker to recover the original plaintext password \cite{shay2010encountering}. To further bolster security, the practice of salting can be employed, wherein random characters are appended to both the beginning and end of the password prior to hashing. This salting technique effectively mitigates the risk of rainbow table attacks and precomputation attacks that target the same hashing function \cite{sriramya2015providing}.

\textit{\textbf{Multi-Factor Authentication.}}
Multi-factor authentication serves as an effective defense strategy that necessitates users to provide additional authentication information in conjunction with their password, thereby ensuring the integrity of the system. This technology is primarily employed within the financial sector. Standard methods of multi-factor authentication include SMS verification codes, hardware tokens, and biometric measures, such as fingerprints and facial recognition. These methods significantly enhance the accuracy and reliability of the authentication process. Even in the event that an attacker acquires a user's password, the requirement for additional authentication information complicates unauthorized login attempts. Implementing multi-factor authentication technology strengthens system security, particularly in applications that handle sensitive data or engage in financial transactions, thereby playing a crucial role in safeguarding such information.

\textit{\textbf{Security Education and Training.}}
Enhancing users' security awareness is critical in mitigating password-cracking attacks and should not be overlooked. To ensure the security of enterprise systems, organizations should implement comprehensive and regular security education and training programs. These programs should cover a range of topics, including the creation of effective passwords, the identification of phishing attacks, and the appropriate handling of potential security threats. As passwords represent a vital component of authentication, their security directly impacts the reliability and availability of the entire system. By improving users' knowledge and awareness regarding network security, organizations can substantially reduce the risks associated with password compromise.

\textit{\textbf{Utilization of Password Managers.}}
A password manager is a tool designed to assist users in securely storing and managing their passwords, thereby ensuring the security and validity of these credentials. The compromise of passwords may lead not only to financial losses for users but also to significant security risks. By utilizing a password manager, users can generate a series of complex, randomly generated passwords tailored for various services. In the event that a password modification is necessary, users only need to input a specific password to execute updates or deletions. As the password manager retains the passwords, users are relieved from the burden of memorizing numerous complex passwords.
Furthermore, to discourage the reuse of identical passwords, each password can be encrypted, enabling users to retrieve the correct information upon decryption. This approach enhances the security of passwords and diminishes the risks associated with their reuse. Additionally, it mitigates the need for extensive specialized knowledge among administrators and key managers. It is essential to note that the password manager itself may be a potential target for attacks; therefore, careful consideration of its security and reliability is paramount when selecting such a tool.

\textit{\textbf{Operating System and Security Software Updates.}}
Regular updates to the operating system and security software are vital in preventing attackers from exploiting known vulnerabilities to gain access to user passwords. These software updates typically include security patches designed to rectify identified security flaws. Users should maintain their operating system and security software in an up-to-date state to minimize the likelihood of successful attacks.

\textit{\textbf{Secure Communications and Encryption.}}
Employing encrypted communication, such as SSL or TLS, is essential for ensuring passwords are not intercepted or compromised during network transmission. Organizations and businesses should prioritize the security of their network connections while employing encryption to safeguard the integrity and confidentiality of user data.

\subsection{Application of Machine Learning to Password Security}

The integration of machine learning into password security is exhibiting a progressively significant trend. The training of machine learning models, coupled with data analysis, provides researchers and security professionals with an enhanced understanding of web security characteristics, improving the efficacy of defense strategies. The utility of machine learning in the realm of password security has been extensively validated through various applications, as outlined below:

\textit{\textbf{Password Strength Assessment.}}
Machine learning can be employed to evaluate the robustness of passwords, enabling users and organizations to make more informed decisions during the creation of such passwords. By analyzing a vast corpus of known passwords, machine learning models can discern patterns associated with weak passwords and predict the strength of newly generated passwords \cite{melicher2016fast}. This methodology aids users in circumventing easily compromised passwords, thereby bolstering account security.

\textit{\textbf{Password Cracking Predictions.}}
Through the utilization of machine learning, security experts can anticipate the strategies and passwords that attackers are likely to target. By training and analyzing models based on a substantial dataset of cracking incidents, researchers can expose the tactics and behavioral traits exhibited by attackers \cite{weir2009password}. Machine learning algorithms facilitate the extraction of all pertinent parameters required for classifying various datasets associated with attacks. This intelligence can be leveraged to devise targeted defense strategies, significantly diminishing the likelihood of successful attacks.

\textit{\textbf{Defense Strategy Optimization.}}
Machine learning can optimize existing password defense strategies, thereby enhancing security and reliability. The application of machine learning algorithms within the field of cryptography is extensive \cite{kelley2012guess}. These algorithms can serve as tools for assessing the effectiveness of multi-factor authentication methods, thus assisting organizations in selecting the authentication protocols that align with their requirements. Furthermore, machine learning can identify and remediate vulnerabilities within security configurations, such as devices operating with default passwords, ultimately improving the overall security posture.

\textit{\textbf{User Behavior Analysis.}}
Machine learning can analyze user behavior to assist organizations in detecting abnormal login attempts and potential security threats. By establishing patterns consistent with normal login behavior, machine learning models can alert users when these patterns are deviated from. This approach effectively identifies compromised accounts and other security concerns promptly, thereby enabling timely defensive actions.

\textit{\textbf{Automated Security Response.}}
The efficiency and consistency of automated security response processes can be significantly enhanced through the application of machine learning. This technology has found various applications within the information security domain. When abnormal login behavior or other suspicious activities are identified, the machine learning system can automatically execute a series of predefined responses, which may include account lockout, notification of administrators, or activation of multi-factor authentication. The implementation of automated response technologies can substantially reduce dependence on human intervention, thereby improving the speed and efficacy of security incident management.

\textit{\textbf{Intelligent Password Generator.}}
Machine learning can be utilized to create intelligent password generators that produce robust and memorable passwords tailored to user requirements and security specifications \cite{li2014emperor}. By training a model to comprehend the structure of human language and lexical patterns, such a generator is capable of crafting passwords that satisfy specific strength criteria while preserving an element of readability. This methodology enhances password security and mitigates the likelihood of users forgetting their passwords.

\textit{\textbf{Password Leakage Monitoring.}}
The application of machine learning allows for real-time monitoring of password leaks on the Internet, providing timely alerts to users so that necessary precautions may be implemented. Moreover, machine learning can be leveraged to analyze user access logs to detect potential attack patterns and offer appropriate recommendations for countermeasures. By employing trained models to scrutinize diverse data sources, including hacker forums, darknet markets, and social media, machine learning can effectively identify compromised passwords and associated information, thereby enhancing data security. This approach enables users to take proactive measures to minimize potential losses prior to the theft of their passwords.

\section{DATA COLLECTION AND PREPROCESSING}
\label{sec::data_collection}
Our study employs real user data publicly leaked from networks to analyze the password database. We collected publicly leaked user information datasets from over 20 websites, including notable platforms such as JD, Taobao, and CSDN, yielding a total user data volume exceeding one billion records \cite{bonneau2012science} \cite{li2014large}. These datasets have been compromised through various methods, including internal leaks by website administrators and external breaches perpetrated by malicious actors who exploit vulnerabilities in social software, email systems, and institutional infrastructures. Such breaches involve using social engineering or weak password protocols to secure unauthorized database access. This section selects 11 representative user datasets based on website types, categorized into lifestyle shopping, social forums, and video entertainment, with detailed sources in Table \ref{tab:web_source}.

\begin{table}[ht]
\small
\centering
\caption{Website Data Sources.}
\vspace{-1ex}
\label{tab:web_source}
\begin{tabular}{>{\centering\arraybackslash}m{2.8cm} >{\centering\arraybackslash}m{1.4cm} >{\raggedright\arraybackslash}m{3.6cm}} 
  \toprule
  \textbf{Type} & \textbf{Volume} & \textbf{Source}\\
  \midrule
  Lifestyle Shopping & 27565970 & Taobao, JD, 12306, Dodonew\\
  \midrule
  Social Forum & 49740893 & Renren, Mop, CSDN, Tianya\\
  \midrule
  Video Entertainment & 36518237 & Duowan, 178, 7k7k\\
  \bottomrule
\end{tabular}
\end{table}

In the investigation of Chinese password databases, six representative user datasets have been selected. These datasets encompass individuals' personal information and associated passwords, totaling 62440436 records. A comprehensive overview of the user datasets is presented in Table \ref{tab:user_source}.

\begin{table}[ht]
\small
\centering
\caption{User Dataset Details.}
\vspace{-1ex}
\label{tab:user_source}
\begin{tabular}{>{\centering\arraybackslash}m{1.2cm} >{\centering\arraybackslash}m{1.4cm} >{\centering\arraybackslash}m{1.2cm} >{\centering\arraybackslash}m{1.2cm} >{\centering\arraybackslash}m{1.4cm} >{\raggedright\arraybackslash}m{5cm}} 
  \toprule
  \textbf{Number} & \textbf{Source} & \textbf{Type} & \textbf{ Volume} & \textbf{Password} & \textbf{Incorporate}\\
  \midrule
  1	& Dodonew	& Shopping	& 16272439	& 10132790	& Serial number, E-mail address, Password\\
  \midrule
  2	& Renren	& Forum	& 4734863	& 2822707	& E-mail, Password\\
  \midrule
  3	& CSDN	& Forum & 4508909	& 2912593 & Email, Username, Password\\
  \midrule
  4	& Mop	& Forum	& 8731286	& 5337012 &	E-mail, Password\\
  \midrule
  5	& 178	& Game	& 9072966	& 3462270	& Serial number, Username, Password\\
  \midrule
  6	& 7k7k	& Game	& 19119973	& 4930689	& E-mail, Password\\
  \bottomrule
\end{tabular}
\end{table}

Data processing entails a multifaceted procedure aimed at extracting valuable insights and knowledge from raw data, necessitating thorough analysis and mining for enhanced comprehension and application. This process generally encompasses several stages, including data cleaning, transformation, feature extraction, and dimensionality reduction. These stages are intrinsically linked to the examination of data types and structures, as well as strategies for utilizing existing databases effectively for processing. In this paper, a range of data processing is employed to augment both the efficiency and accuracy of the data processing:

\begin{enumerate}
\item Data De-duplication: Duplicate passwords within the dataset are eliminated to ensure that each password entry is distinct and independent.
\item Outlier Handling: Passwords that are excessively long or short are discarded, retaining only those within a prescribed length range (e.g., 4-20 characters). This procedure serves to filter out anomalous data entries that may result from errors or systemic generation.
\item Illegal Character Processing: A review is conducted to identify and remove illegal characters from the user dataset, thereby ensuring data integrity and accuracy. Illegal characters may include unprintable ASCII characters and control characters. Their removal is critical to prevent distortion in subsequent analysis outcomes.
\item Data Desensitization: To safeguard user privacy, all sensitive information, including usernames and email addresses, is desensitized. This approach focuses solely on the password itself, without disclosing any personal identifiers.
\end{enumerate}

\section{DATA CHARACTERIZATION}
\label{sec::characterization}

This section will thoroughly analyze the characteristics of password data sources, including analysis of familiar patterns, password length, and character composition, to uncover users' password selection habits and vulnerabilities across different websites \cite{florencio2007large}. By comparing the strength and characteristics of passwords from each data source, this study aims to provide tailored password security recommendations and protective measures for enterprises and individuals.

\subsection{Familiar Patterns}

A ranking of the top ten most prevalent passwords has been compiled by sorting them based on their frequency of occurrence in the user datasets, with results presented in Table \ref{tab:popular_password}. The proportion of these top ten passwords ranges from 3.30\% (Shopping 1) to 10.76\% (Forum 1). This variance suggests that distinct website types and user demographics exhibit differing preferences in password selection \cite{mittal2019demographic}. Based on the analyzed password patterns, categories have been delineated as follows: sequential numbers, repeated characters, keyboard patterns, simple fragment combinations, and semantically significant passwords.

\begin{enumerate}
 \item Sequential Numbers and Repeated Characters: Common passwords, such as "123456", "12345678", and "111111", reveal a notable inclination among users to incorporate consecutive numbers and repeated characters into their passwords. Such conventions make these passwords easier to compromise due to their predictable nature.
 \item Keyboard Patterns: Certain popular passwords, including "w2w2w2" and "147258369," highlight a tendency for users to utilize sequences of characters situated on the keyboard as passwords. This makes it susceptible to cracking attacks predicated on recognizable keyboard patterns.
 \item Simple Words and Alphanumeric Combinations: Passwords such as "a123456", "123456a", and "qq66666" indicate that users favor simple words combined with alphanumeric elements. Although these options present a higher complexity than purely numeric passwords, they nonetheless lack adequate strength to withstand brute-force attacks.
 \item Special Semantic Passwords: Passwords like "5201314" and "1314520", which carry special semantic meaning, suggest that users might select personally significant terms. Examples such as "qiulaobai", "111222tianya", and "dearbook" may connect to specific websites or cultural references. While these passwords could be more challenging to guess, they remain vulnerable to targeted attacks if an assailant conducts a thorough analysis of the relevant cultural context or website content \cite{mazurek2013measuring}.
\end{enumerate}

\begin{table}[ht]
  \small
  \centering
  \caption{Popular Passwords.}
  \vspace{-1ex}
  \label{tab:popular_password}
  \begin{tabular}{>{\centering\arraybackslash}m{1.6cm} >{\centering\arraybackslash}m{1.6cm} >{\centering\arraybackslash}m{1.6cm} >{\centering\arraybackslash}m{1.6cm} >{\centering\arraybackslash}m{1.6cm} >{\centering\arraybackslash}m{1.6cm}  >{\centering\arraybackslash}m{1.6cm}} 
    \toprule
    \textbf{Ranking} & \textbf{Game1} & \textbf{Game2} & \textbf{Forum 1} & \textbf{Shopping 1} & \textbf{Forum 2} & \textbf{Forum 3}\\
    \midrule
    1	& 123456	& 123456	& 123456789	& 123456	& 123456	& 123456\\
    \midrule
    2	& 111111	& 111111	& 12345678	& a123456	& 123456789	& 123456789\\
    \midrule
    3	& 123456789	& zz12369	& 11111111	& 123456789	& 111111	& 111111\\
    \midrule
    4	& 123123	& qiulaobai	& dearbook	& 111111	& 123123	& 123123\\
    \midrule
    5	& 111222tianya	& 123456aa	& 00000000	& 5201314	& 5201314	& 5201314\\
    \midrule
    6	& 5201314	& wmsxie123	& 123123123	& 123123	& 789456	& 12345\\
    \midrule
    7	& 123321	& 123123	& 1234567890	& a321654	& 666666	& 12345678\\
    \midrule
    8	& 12345678	& 000000	& 88888888	& 12345	& 123321	& 123321\\
    \midrule
    9	& 666666	& qq66666	& 111111111	& 000000	& 1314520	& 1314520\\
    \midrule
    10	& 7758521	& w2w2w2	& 147258369	& 123456a	& 12345678	& 1234567\\
    \midrule
    \% of top-10 & 6.55\%	& 8.74\%	& 10.76\%	& 3.30\%	& 7.44\%	& 6.68\%\\
    \bottomrule
  \end{tabular}
  \end{table}

\subsection{Password Length}

In the analysis of password databases, the length of passwords serves as a significant indicator. It represents one of the most discernible characteristics of password data distribution, as it directly influences the security of these passwords. Generally, passwords with greater lengths exhibit a heightened resistance to brute-force attack methods. However, users tend to select shorter or simpler passwords for ease of memorization during the password setup process. The length of a password is closely correlated with the password-setting policies enacted by various websites. Different types of websites implement distinct password-setting strategies, which in turn affect users' choices when establishing their passwords. Many websites impose specific password policies, mandating that users create passwords of greater length and complexity.

\begin{figure}[htbp]
 \centering
 \includegraphics[width=0.5\textwidth]{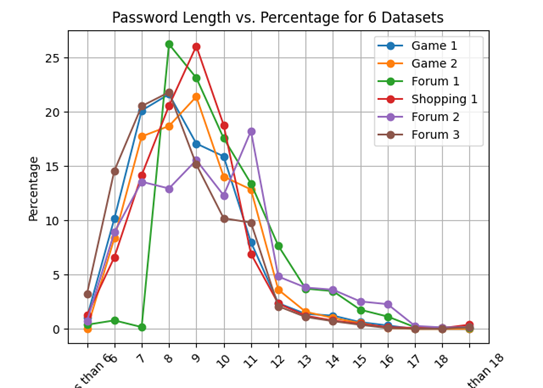}
 \caption{Distribution of Password Lengths for Six Typical Datasets.}
 \label{fig:password_length}
 \vspace{-2ex}
\end{figure}

This paper presents an analysis of password length data across six representative datasets, the results of which are illustrated in Fig. \ref{fig:password_length}. The findings are summarized as follows:

\begin{itemize}
 \item In the shopping dataset, there exists a notable prevalence of passwords with lengths of 8 and 9 characters, at 20.57\% and 26.07\%, respectively, culminating in a combined total of 46.64\%. This trend suggests that users are inclined to utilize medium-length passwords on shopping websites, potentially due to the significant security requirements established by these platforms. Furthermore, the proportion of passwords with lengths shorter than six characters in the shopping dataset is notably low, at merely 1.26\%, further reflecting the password length requirements implemented by shopping websites.
 \item In the game dataset, a higher percentage of mnemonics with lengths of 7 and 8 characters is observed in Game 1, at 20.13\% and 21.63\% respectively, which totals 41.76\%. Similarly, Game 2 shows a preference for passwords of lengths 7 and 8, with percentages of 17.74\% and 18.70\%, totaling 36.44\%. This pattern indicates a tendency among users to favor shorter passwords on gaming websites, likely attributable to the relatively lower password security requirements on these platforms and a preference for convenience in memorization. Notably, the percentage of passwords with a length of 6 in the Game 1 dataset reaches 10.16\%, which is significantly higher than that in the other datasets.
 \item The distribution of password lengths within the forum dataset displays a more decentralized pattern. In the Forum 1 dataset, the highest proportion of passwords corresponds to a length of 8 characters, reaching 26.24\%. In Forum 2, the percentage of passwords with a length of 7 is recorded at 13.57\%. At the same time, those with lengths between 7 and 11 characters also exceed 10\%, with the 11-character passwords accounting for 18.24\%, denoting a relatively long password. Conversely, Forum 3 indicates that the percentages of passwords with lengths of 6, 7, 8, and 9 are relatively high, at 14.58\%, 20.53\%, 21.81\%, and 15.16\%, respectively.
\end{itemize}

In this paper, we analyze and compare the lengths of passwords across three distinct datasets: shopping, gaming, and forum platforms. The findings reveal several noteworthy characteristics:

\begin{enumerate}
 \item Distribution of Password Length: The datasets for shopping and gaming exhibit a higher proportion of passwords ranging from 8 to 10 characters. Conversely, the forum dataset predominantly features longer passwords. This discrepancy suggests variations in password strategies among different types of websites. Users of shopping and gaming platforms tend to favor shorter passwords, whereas users of forum websites are more inclined to opt for longer passwords.
 \item Website Password Strategy: Our analysis indicates that the password strategies employed by shopping and gaming websites are relatively lenient, permitting users to select shorter passwords. This approach enhances user experience and alleviates the cognitive load associated with memorizing passwords. However, it also significantly compromises password security. In contrast, forum websites impose more stringent requirements, necessitating users to establish longer passwords to bolster security. It may reflect a greater emphasis on privacy and information security within forum platforms.
 \item Password Security: From a security perspective, longer passwords demonstrate increased resilience against brute-force attacks. For users of shopping and gaming websites, reliance on shorter passwords may introduce heightened security risks. Consequently, website administrators should adopt an appropriate password strategy that balances user experience with security considerations. Users should likewise acknowledge the critical importance of password security and strive to select passwords of moderate length and complexity to enhance protection.
\end{enumerate}

\subsection{Password Character Composition}

The strength and security of a password are significantly influenced by its character composition \cite{cai2016collective}. An in-depth analysis of password character composition provides insight into the types of characters utilized by users when creating passwords. We have conducted an assessment of the frequency and distribution of numerical digits, lowercase letters, uppercase letters, and special characters within the password dataset.

This paper investigates the password character composition across six representative datasets, with results illustrated in Fig. \ref{fig:password_distribution}. The character types within the password dataset are categorized into four distinct types: digits (D), uppercase letters (U), lowercase letters (L), and symbols (S). The composition of these character types is ranked in descending order of prevalence. Based on the analysis of the dataset mentioned above, the following conclusions can be derived:

\begin{figure}[htbp]
 \centering
 \includegraphics[width=0.6\textwidth]{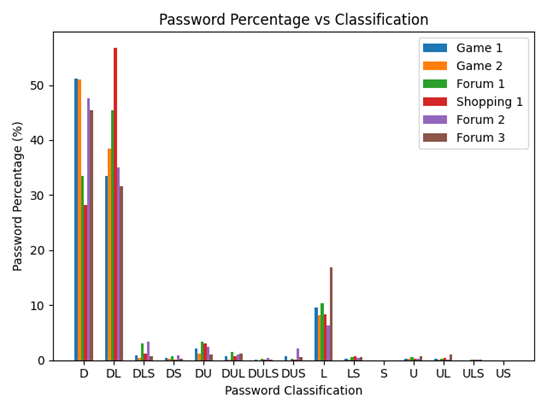}
 \caption{Distribution of Password Character Composition for Six Typical Datasets.}
 \label{fig:password_distribution}
 \vspace{-2ex}
\end{figure}

\begin{itemize}
 \item The proportion of passwords predominantly comprising a combination of numbers and lowercase letters (DL) is notably high across all datasets, particularly within the Shopping 1 dataset, where it reaches 56.81\%. This trend suggests that users favor this combination for password creation, likely due to its relative ease of memorization.
 \item The occurrence of numeric (D) passwords utilized independently is significantly elevated across various datasets. For instance, the metrics for the Game 1 and Game 2 datasets reflect proportions of 51.23\% and 50.97\%, respectively. It indicates a user preference for operational convenience over enhanced security when formulating passwords, thus highlighting the level of concern users exhibit regarding password security.
 \item The prevalence of passwords incorporating numbers, uppercase letters, lowercase letters, and special characters (DULS) remains exceedingly low across all datasets. The underutilization of more secure character combinations suggests a general lack of emphasis on password security.
 \item The distribution of password character compositions demonstrates variability across different types of datasets, including shopping, gaming, and forums. In the Shopping 1 dataset, the percentage of numeric (D) passwords stands at 28.17\%, whereas in the Game 1 and Game 2 datasets, this figure escalates to 51.23\% and 50.97\%, respectively. This discrepancy may be attributable to varying password security policy requirements across distinct online platforms.
\end{itemize}

\subsection{Summary}
After thorough analysis and evaluation above, we determined that these datasets exhibit the following main characteristics:
\begin{enumerate}
 \item Extensive Data Volume: The password database utilized in this study encompasses a vast amount of data, comprising tens of millions of user password records, which serves to reflect users' password selection behaviors.
 \item Diversity of Password Types: The dataset includes a variety of password formats, such as numeric-only, alphabetic-only, mixed alphanumeric, and those incorporating special characters, thereby elucidating users' password selection tendencies across different scenarios. 
 \item Multiple Data Sources: The password data is sourced from various websites, applications, and industries. This multiplicity of sources contributes to a comprehensive understanding of the characteristics inherent in user password selections. 
 \item Temporal Relevance: The dataset spans a nearly decade-long timeframe, facilitating an examination of trends in the evolution of password strategies and user behaviors.
 \item Representative Sampling: Eleven representative data sources were chosen for analysis to elucidate users' password selection habits and vulnerabilities across different categories of websites.
 \item Sensitive Information Management: Although the user datasets are publicly accessible and have been utilized in research concerning password security, they contain sensitive information regarding individual website users, including email addresses and passwords \cite{cai2016collective}. To mitigate the security risks associated with the characteristics of the leaked data sources, identifying information has been removed in subsequent analyses and replaced with numerical designations (e.g., Forum1, Game1, Shopping1). 
\end{enumerate}

\subsection{Common Attack Patterns and Threats}

Based on the password characterization outlined above, it is apparent that the user datasets may be vulnerable to the following prevalent attack patterns:

\begin{enumerate}
 \item Brute-force cracking: A significant number of users opt for short passwords and simplistic sequential numbers, repeated characters, or basic alphanumeric combinations. These choices render brute-force cracking attacks a viable threat, particularly within the Game 1, Game 2, and Forum 1 datasets, where short passwords are more prevalent.
 \item Dictionary attack: Attackers can exploit a precompiled dictionary (containing familiar words, phrases, and combinations) to compromise accounts using simplistic word and phrase combinations as passwords. Such passwords are frequently encountered in the Shopping 1, Forum 2, and Forum 3 datasets, thereby amplifying the risk of dictionary attacks in these instances.
 \item Social engineering: Attackers may leverage personal information, including birthdays and names, to deduce passwords associated with individuals. Additionally, passwords connected to specific numbers (e.g., "5201314" and "1314520") and those relevant to particular websites or cultural references are vulnerable to social engineering tactics. These password types are more common in the Shopping 1 dataset, thus increasing the likelihood of social engineering threats.
 \item Keyboard pattern attacks: Attackers may focus on accounts that employ sequential character patterns derived from the keyboard as passwords. Such passwords are predominantly found in the Game 1, Game 2, and Forum 1 datasets, suggesting that keyboard pattern attacks may pose a significant risk to these datasets.
 \item Targeted attacks: Attackers may engage in focused efforts to guess passwords based on factors such as the distinct characteristics of the target site and user demographics \cite{hitaj2019passgan}. For instance, attempts may be made to guess semantic passwords that are culturally relevant to specific websites. These types of passwords are prevalent in the Forum 2 and Forum 3 datasets, indicating that targeted attacks may represent a considerable threat to these datasets.
\end{enumerate}

\section{MACHINE LEARNING BASED CLASSIFIER FOR STRONG AND WEAK PASSWORDS}
\label{sec::classifier}

This study employs a structured methodology to develop a machine learning-based classifier for identifying strong and weak passwords:

\begin{enumerate}
 \item Data Collection and Preprocessing: The initial phase involves gathering a substantial volume of password data from a publicly available password database. This data is subsequently subjected to a rigorous cleaning process, where duplicates are removed, and preprocessing is conducted to ensure compatibility with subsequent feature extraction and model training requirements.
 \item Feature Extraction: For each password, an array of features is extracted, including password length, character type (encompassing numeric, uppercase, and lowercase letters, as well as special characters), frequency of character repetition, the most extended consecutive sequence of identical characters, and the degree of character type variation. These features enable machine learning models to comprehend the structural characteristics and security level of each password.
 \item Data Labeling: Each password is assigned a strength classification, such as "strong" or "weak," based on its intrinsic characteristics. This classification serves as the target variable during the supervised learning phase, guiding the model's training and optimization.
 \item Dataset Division: The preprocessed dataset is partitioned into three segments: the training set, the validation set, and the test set. The training set is utilized for model training; the validation set is employed for hyperparameter adjustment and optimal model selection, and the test set is reserved for evaluating the final model's performance.
 \item Model Selection and Training: Multiple machine learning algorithms, including support vector machines, logistic regression, neural networks, decision trees, random forests, and stacked models, are evaluated. The training set is utilized to train these algorithms, and suitable optimization are selected based on the specific characteristics of both the algorithms and the problem at hand.
 \item Model Validation and Hyperparameter Tuning: The performance of each model is rigorously assessed using the validation set, with key evaluation metrics such as accuracy, recall, and F1 score being calculated. During the training phase, model performance is refined by adjusting hyperparameters. This optimization process can occur concurrently with model validation, employing techniques such as cross-validation or utilizing validation sets to gauge model performance across various hyperparameter configurations.
 \item Model Evaluation: Ultimately, the performance of the selected model is assessed using the test set. The evaluation metrics of accuracy, recall, and F1 score are computed based on the model's predictions compared to the actual labels.
\end{enumerate}

\subsection{Pre-preparation}

\subsubsection{Data Collection and Preprocessing}
The dataset utilizes password information compiled from the analysis of a public password database as described in Section \ref{sec::data_collection}, specifically drawing from a user dataset that was leaked from the website CSDN.
To facilitate effective model learning from clean and representative data,  we extensively cleaned and preprocessed the collected password information. This process entails the removal of duplicate entries to maintain data uniqueness and the addressing of any anomalies, including the exclusion of passwords that are either excessively short or lengthy. The relevant preprocessing measures have been discussed previously in the analysis of the public password database in Section \ref{sec::data_collection}.

\subsubsection{Feature Engineering}

Feature engineering represents a critical phase in the machine learning process, involving the extraction, construction, and selection of features from raw data that enhance model performance. In the context of classifying strong and weak passwords, the objective of feature extraction and processing is to derive meaningful insights from the raw passwords, thereby enabling the machine learning model to comprehend and learn from this data more effectively. The subsequent section delineates the feature extraction process pertinent to this research \cite{ur2012does}:

\begin{enumerate}
 \item Length: The length of the password is initially computed, serving as a pivotal feature because longer passwords are generally more challenging to compromise than their shorter counterparts. This feature directly indicates the total number of characters present in the password.
 \item Character Types: The classification of different character types (including numbers, lowercase letters, uppercase letters, and special characters) within a password constitutes a feature that aids in evaluating the password's complexity. Typically, a password exhibiting multiple character types is more resilient against attacks than one composed of a single character type.
 \item Level of Character Repetition: Assessing the degree of character repetition within a password can assist in evaluating the diversity of characters utilized. A higher level of repetition may render the password more susceptible to compromise.
 \item Maximum Number of Consecutive Identical Characters: This feature reflects the length of the most extended sequence of consecutive identical characters present in the password. A protracted series of identical characters may indicate a weaker password strength.
 \item Degree of Character Type Variation: The enumeration of character type variations in the password serves as a measure of its complexity. Frequent transitions between character types may signify a heightened complexity, thus enhancing the password's resistance to attacks.
\end{enumerate}

The results obtained following the extraction of password features are presented in Table \ref{tab:password_feature}.

\begin{table}[ht]
\small
\centering
\caption{Password Characteristics.}
\vspace{-1ex}
\label{tab:password_feature}
\begin{tabular}{>{\centering\arraybackslash}m{3cm} >{\raggedright\arraybackslash}m{8cm} } 
  \toprule
  \textbf{Feature} & \textbf{Content}\\
  \midrule
  length	& Length of the password\\
  \midrule
  num\_digits	& Number of digits in the password\\
  \midrule
  num\_lowercase	& Number of lowercase letters in the password\\
  \midrule
  num\_uppercase	& Number of capital letters in the password\\
  \midrule
  num\_special\_chars	& Number of special characters (e.g., symbols) in the password\\
  \midrule
  char\_repea	& Number of repeated characters in the password\\
  \midrule
  max\_consecutive\_chars	& Maximum number of consecutive identical characters in the password\\
  \midrule
  char\_type\_changes	& Number of times the character type (numeric, lowercase, uppercase, special characters) in the password has changed\\
  \midrule
  password\_strength	& password strength, which is the target variable, usually expressed as strong (1) or weak (0)\\
  \bottomrule
\end{tabular}
\end{table}

\subsubsection{Labeling}

Labeling, in this context, pertains to the assignment of a label to each password that signifies its strength. This labeling process involves a binary categorization that denotes whether the password is classified as strong (1) or weak (0). A password qualifies as strong (labeled 1) if its length is equal to or greater than nine characters and encompasses at least three different character types (numbers, lowercase letters, uppercase letters, and special characters). In contrast, a password that does not meet these criteria is denoted as weak (labeled 0). This labeling system will be employed in subsequent machine learning processes to train the model to assess the strength of passwords automatically.

\subsubsection{Data Segmentation}

The complete dataset is partitioned into a training set, validation set, and test set. The training set instructs the model, the validation set performs hyperparameter optimization, and the test set evaluates the model's final performance. This partitioning is conventionally executed in the ratio of 70\% for training, 15\% for validation, and 15\% for testing.

\subsubsection{Feature Scaling}

It is imperative to normalize or standardize the features to mitigate optimization issues arising from diverse ranges of feature values. Standard techniques include Min-Max scaling and Z-score standardization. In this analysis, the normalization method is implemented to scale the feature data. The normalization process modifies the feature data according to Equation \ref{eqa:normalization}:

\begin{equation}\label{eqa:normalization}
 X_{\text{norm}} = \frac{X - \mu}{\sigma}
\end{equation}

In this equation, X represents the original feature data, $\mu$ signifies the average of the feature, and $\sigma$ indicates the standard deviation of the feature. This transformation results in feature data exhibiting a normal distribution characterized by a mean of 0 and a standard deviation of 1.

\subsubsection{Data Balancing}

In the scenario of password strength classification, an imbalance in the distribution of strong and weak passwords may lead to the classifier favoring the category with a greater number of instances during the prediction process. To address this challenge, oversampling and undersampling can be employed to achieve data balance. The dataset utilized in this study exhibits an imbalance concerning the distribution of strong and weak passwords; consequently, the undersampling method is implemented for data balancing purposes. 

Undersampling \cite{batista2004study}, such as RandomUnderSampler, balance the dataset by randomly removing samples from the category with a larger representation. While undersampling minimizes the risk of overfitting, it may also result in the loss of valuable information.

\subsection{Model Training}

\subsubsection{Model Selection}
A variety of machine learning algorithms can be employed, and their performance can be assessed in addressing the classification problem of strong and weak passwords. The algorithms utilized in this study include:

Logistic Regression: Logistic regression serves as a straightforward and easily implemented linear classifier suitable for binary classification tasks \cite{bishop2006pattern}. It provides the probability of predicting an outcome. It is particularly effective for datasets after appropriate feature engineering, primarily when there exists a substantial linear relationship between the features and the target variable.

Support Vector Machine (SVM): A support vector machine functions as a nonlinear classifier that seeks to identify a hyperplane maximizing the margin between positive and negative samples. SVMs are especially beneficial for high-dimensional and nonlinear problems; however, they may exhibit high computational complexity when applied to large datasets.

Neural Networks: Neural networks, recognized for their capability as robust nonlinear classifiers, can effectively extract complex features \cite{kim2023multi}. Simple neural network architectures, such as multilayer perceptrons (MLP), are recommended for strong and weak password classification. Although requiring extended training durations, they produce superior classification outcomes.

Decision Trees: A decision tree establishes a classification model via the recursive partitioning of a dataset \cite{quinlan1986induction}. The model is simple to comprehend and interpret; however, it is susceptible to overfitting. This issue can be alleviated by adjusting hyperparameters such as tree depth and the number of leaf nodes.

Random Forest: Random Forest is an ensemble learning technique grounded in decision trees \cite{breiman2001random}. It enhances classification performance by generating multiple decision trees and amalgamating their predictions through a voting mechanism. This approach offers improved generalization capabilities and mitigates the risk of overfitting.

Stacking: Stacking is an advanced ensemble learning strategy that integrates the predictions of multiple base models (also referred to as base learners) to yield a more precise aggregate prediction \cite{wolpert1992stacked}. This method combines the forecasts of base classifiers by training a secondary classifier, known as a meta-learner or meta-classifier, which learns to optimally derive a composite prediction from the individual predictions of the base classifiers.

\subsubsection{Hyperparameter Adjustment}
The performance of models can be optimized through the fine-tuning of their hyperparameters during the training phase. Hyperparameters represent external configuration parameters in a machine-learning model that are not automatically updated throughout the training process. The selection of hyperparameters has a profound effect on model performance. To identify the most effective combination of hyperparameters, Grid Search is employed in this study \cite{bergstra2012random}. This exhaustive search method evaluates all possible combinations of hyperparameters to determine the optimal configuration. While Grid Search is highly effective within smaller search spaces, an increase in the search space demands significantly greater computational time and resources, necessitating the use of parallel processing on the computing system. The hyperparameters identified via grid search are detailed in Table \ref{tab:parameters}.

\begin{table}[ht]
\small
\centering
\caption{Model Hyperparameters.}
\vspace{-1ex}
\label{tab:parameters}
\begin{tabular}{>{\centering\arraybackslash}m{3cm} >{\centering\arraybackslash}m{7cm} } 
  \toprule
  \textbf{Algorithm Type} & \textbf{Parameters}\\
  \midrule
  SVM	& C: 10, kernel: rbf\\
  \midrule
  Logistic Regression	& C: 1, max\_iter: 100, solver: newton-cg\\
  \midrule
  Neural Networks	& activation: relu, hidden\_layer\_sizes: (100,), max\_iter: 500, solver: adam\\
  \midrule
  Decision Trees	& criterion: gini, max\_depth: None, min\_samples\_leaf: 1, min\_samples\_split: 2\\
  \midrule
  Random Forest	& criterion: entropy, max\_depth: 20, min\_samples\_leaf: 1, min\_samples\_split: 10, n\_estimators: 10\\
  \midrule
  Stacking	& dt\_criterion : gini, dt\_max\_depth: None, svm\_C: 0.1, svm\_gamma : scale, svm\_kernel : linear\\
  \bottomrule
\end{tabular}
\end{table}

\subsubsection{Model Validation}

Model validation is utilized to assess a model's performance on previously unseen data. A validation set, which is a subset of the original dataset, is partitioned to evaluate the model during the training process. Significantly, the validation set is not utilized in the model training, thereby serving as a representative sample of unknown data.

During the model validation phase, the validation set is employed to gauge the model's performance. By comparing the model's predictions with the actual labels present in the validation set, various evaluation metrics, including accuracy, recall, and the F1 score, can be calculated.

\begin{enumerate}
 \item Accuracy: Accuracy is defined as the ratio of the number of correctly predicted instances to the total number of instances \cite{powers2020evaluation}. This metric is particularly relevant when there is a balanced distribution of categories. In cases of category imbalance, accuracy may appear disproportionately high, potentially misrepresenting the model's performance.
 \item Recall: Recall is calculated as the ratio of True Positive cases to the total number of actual positive cases, which includes both True Positives and False Negatives \cite{powers2020evaluation}. This metric assesses the model's capability to identify positive examples, making it particularly significant where the recognition of positive instances, such as anomaly detection and disease diagnosis, is required.
 \item F1 Score: The F1 score represents the harmonic mean of accuracy and recall and is utilized to provide a comprehensive evaluation of the model's performance \cite{powers2020evaluation}. High values in both accuracy and recall will yield a high F1 score. This metric is especially beneficial in scenarios involving category imbalance, as it offers a more accurate representation of model performance in contexts focused on positive instance recognition.
\end{enumerate}

The performance of six models is evaluated utilizing validation datasets, as detailed in Table \ref{tab:validation_score}.

\begin{table}[ht]
\small
\centering
\caption{Accuracy, Recall, and F1 Score for Six Models.}
\vspace{-1ex}
\label{tab:validation_score}
\begin{tabular}{>{\centering\arraybackslash}m{3cm} >{\centering\arraybackslash}m{1.2cm} >{\centering\arraybackslash}m{1.2cm} >{\centering\arraybackslash}m{1.2cm}} 
  \toprule
  \textbf{Algorithm Type} & \textbf{Accuracy}	& \textbf{Recall}	& \textbf{F1 Score}\\
  \midrule
  SVM	& 0.9891	& 0.9952	& 0.8846\\
  \midrule
  Logistic Regression	& 0.9329	& 0.8455	& 0.5133\\
  \midrule
  Neural Networks	& 0.9941	& 0.9984	& 0.9337\\
  \midrule
  Decision Trees	& 0.9998	& 1.0000	& 0.9976\\
  \midrule
  Random Forest	& 0.9955	& 1.0000	& 0.9494\\
  \midrule
  Stacking	& 0.9998	& 1.0000	& 0.9976\\
  \bottomrule
\end{tabular}
\end{table}

In this analysis, six machine learning algorithms were employed to assess their performance in the classification of strong and weak passwords. A detailed comparison of the accuracy, recall, and F1 scores of each model yields the following conclusions:

\begin{itemize}
 \item The Decision Tree (DT) model exhibited superior performance among all evaluated models, achieving an accuracy of 0.9998, a recall of 1.0000, and an F1 score of 0.9976. These results indicate the model's robust accuracy and reliability in the classification task.
 \item The Random Forest model also demonstrated commendable performance with an accuracy of 0.9955, a recall of 1.0000, and an F1 score of 0.9494. As an ensemble algorithm, Random Forest capitalizes on the strengths of multiple decision trees, enhancing classification capabilities through a combination of their predictions.
 \item The Stacking model, another ensemble method, integrates the outputs of various base models as new features for a meta-model. It achieved an accuracy of 0.9998, a recall of 1.0000, and an F1 score of 0.9976, which is comparable to the performance of the Decision Tree model.
\end{itemize}

Through a comprehensive analysis, it is evident that the Decision Tree, Random Forest, and Stacking models performed exceptionally well in this classification task, as evidenced by high accuracy, recall, and F1 scores. In contrast, the performance of Neural Networks and Support Vector Machines was comparatively lesser, while Logistic Regression revealed relatively poorer results.

\subsection{Model Evaluation}

Model evaluation represents the final phase following the training and validation of a model, with the objective of assessing the model's performance on previously unseen data \cite{de2014very}. In this stage, we evaluate the selected model utilizing a test set and compute various evaluation metrics, including accuracy, recall, and F1 score, to establish a benchmark for the model's anticipated performance in real-world applications.
We evaluate model performance based on the test set for the six models, as detailed in Table \ref{tab:test_score}.

\begin{table}[ht]
\small
\centering
\caption{Accuracy, Recall, and F1 Score for Six Models.}
\vspace{-1ex}
\label{tab:test_score}
\begin{tabular}{>{\centering\arraybackslash}m{3cm} >{\centering\arraybackslash}m{1.2cm} >{\centering\arraybackslash}m{1.2cm} >{\centering\arraybackslash}m{1.2cm}} 
  \toprule
  \textbf{Algorithm Type} & \textbf{Accuracy}	& \textbf{Recall}	& \textbf{F1 Score}\\
  \midrule
  SVM	& 0.9903	& 0.9968	& 0.8940\\
  \midrule
  Logistic Regression	& 0.9326	& 0.8576	& 0.5118\\
  \midrule
  Neural Networks	& 0.9937	& 1.0000	& 0.9293\\
  \midrule
  Decision Trees	& 0.9998	& 1.0000	& 0.9976\\
  \midrule
  Random Forest	& 0.9950	& 1.0000	& 0.9428\\
  \midrule
  Stacking	& 0.9998	& 1.0000	& 0.9976\\
  \bottomrule
\end{tabular}
\end{table}

The ensuing analysis provides a detailed examination of the performance of the six models based on the results from the evaluation of the test set: 

\begin{itemize}
 \item The Support Vector Machine demonstrates superior performance across the metrics of accuracy, recall, and F1 value, particularly in the recall, which reaches 0.9968, indicating its efficacy in incorrect classification.
 \item Logistic regression exhibits overall modest performance in terms of accuracy, accompanied by low recall and F1 values; this suggests deficiencies in its ability to classify strong and weak passwords effectively.
 \item The neural network performs commendably in terms of both accuracy and recall; however, the F1 value is somewhat lower, likely owing to occasional inaccuracies in predicting specific categories.
 \item The decision tree excels on all evaluation metrics, achieving an accuracy of 0.9998, a recall of 1.0000, and an F1 value of 0.9976, thereby indicating its exceptional capacity for strong and weak password classification.
 \item Random forests also exhibit outstanding performance in accuracy and recall; nevertheless, the F1 value is slightly diminished, potentially due to inaccuracies in predicting certain categories under specific circumstances.
 \item The stacking model shows considerable efficacy in accuracy, recall, and F1 value, rivaling that of decision trees. The stacking model effectively amalgamates the strengths of multiple base models, thereby offering a robust solution for the strong and weak password classification tasks.
\end{itemize}

In summary, the evaluation results on the test set indicate that Decision Trees and Stacking Models are well-suited for strong and weak password classification, exhibiting commendable performance across accuracy, recall, and F1 value metrics. While Neural Networks and Support Vector Machines perform admirably in specific evaluation metrics, they are slightly less effective than Decision Trees and Stacking Models in terms of F1 values. Conversely, Logistic Regression and Random Forests are comparatively weaker and less appropriate for the classification of strong and weak passwords.

\section{DEFENCE STRATEGIES RECOMMENDATION}
\label{sec::defense_strategy}

It is imperative to implement a variety of effective password policies and management strategies to enhance password security \cite{chiasson2008influencing}. The following recommendations are put forth:

\begin{enumerate}
 \item Establishment of Complexity Requirements: It is essential to ensure that the user-generated password includes a combination of uppercase letters, lowercase letters, numbers, and special characters. This multifaceted approach significantly increases the difficulty of password cracking, thereby enhancing overall security.
 \item Implementation of Minimum Length Standards: A minimum length requirement should be established for passwords, recommending a baseline of at least eight characters. Generally, longer passwords tend to be more challenging to guess or compromise.
 \item Encouragement of Periodic Password Changes: Users should be encouraged or mandated to change their passwords at regular intervals, such as every three months. This practice serves to mitigate the risk of password theft.
 \item Prohibition of Common Words and Simple Patterns: Users should be restricted from utilizing common words, consecutive numbers, or easily identifiable patterns (e.g., "123456," or "admin") as passwords. Such choices are particularly vulnerable to dictionary and brute-force attacks.
 \item Limitation of Login Attempts: The system should restrict the number of consecutive incorrect password entries allowed within a defined time frame. For instance, account access could be temporarily suspended after five unsuccessful attempts.
 \item Enforcement of Password History Restrictions: Users should be prohibited from reusing passwords that have been utilized in previous accounts. For example, an individual should be barred from selecting any of the last five used passwords during the update process.
 \item Integration of Secondary Authentication: The implementation of two-factor authentication, such as SMS-based CAPTCHAs, hardware tokens, or biometric verification, is recommended to fortify account security further. This additional layer ensures that even if a password is compromised, unauthorized access is impeded without the second factor.
 \item Conducting Audits and Monitoring: Ongoing assessments of password strength, along with monitoring of account activities, are crucial for the timely detection of suspicious behavior. For example, unusual login locations and frequent password resets should be flagged for review.
 \item Enhancement of Personal Security Awareness: It is vital to differentiate between commonly used personal passwords and those utilized in a professional context. Users should be advised to refrain from incorporating personal information into their work-related passwords.
\end{enumerate}

\subsection{Multi-Factor Authentication}

Multi-Factor Authentication (MFA) represents a robust security authentication method that necessitates users to validate their identity by employing two or more distinct types of authentication factors \cite{kim2011method}. The underlying principle of Multi-Factor Authentication is predicated upon three fundamental elements: a knowledge factor, which comprises information known to the user, such as a password or Personal Identification Number (PIN); a possession factor, encompassing an item owned by the user, such as a smartcard or security token; and a biometric factor, which utilizes the user's biometric identifiers, such as a fingerprint or facial recognition \cite{ometov2018multi}. The implementation of multi-factor authentication in critical systems significantly enhances security. In environments that handle financial data, sensitive information, or critical infrastructure, MFA can effectively thwart attackers from gaining access by compromising a single authentication factor, such as a password. Consequently, it is highly advisable to deploy multi-factor authentication across critical systems to bolster overall security. Below are some widely utilized multi-factor authentication methods:

\begin{enumerate}
 \item SMS Verification: Upon entering the username and password, the system transmits a Short Message Service (SMS) containing a verification code to the user's pre-registered mobile phone number. The user should enter the received code to complete the authentication process. This method integrates a knowledge factor (password) and a possession factor (mobile phone).
 \item Fingerprint Identification: Users are required to authenticate through a fingerprint scanner. This method combines a biometric factor (fingerprint) and a knowledge factor (password).
 \item Dynamic Tokens: The user should associate a physical device, such as a security token, with their account. This device generates a dynamic verification code that the user should enter to finalize the authentication process. This methodology integrates a possession factor (security token) and a knowledge factor (password).
 \item Application Authentication: Users can generate dynamic authentication codes using specialized applications, such as Google Authenticator or Microsoft Authenticator. This procedure combines a possession factor (smartphone) and a knowledge factor (password).
\end{enumerate}

\subsection{Regular Audits and Updates}

Given the continuous evolution of network security threats, reliance on a static password policy is insufficient for maintaining long-term security. Regular audits and updates to password policies are imperative to safeguard system and data integrity. Organizations and individuals alike should conduct periodic evaluations and updates of their password policies to address the ever-changing landscape of security threats effectively. Regular audits facilitate the identification of potential security vulnerabilities, allowing for timely remediation of issues that could result in data breaches or system compromises. By consistently auditing and revising password policies, entities can enhance their protection against cyber threats. Adapting to emerging security challenges is crucial for ensuring the protection of systems and data. The following recommendations are suggested:

\begin{enumerate}
 \item Regular Evaluation of Password Policies: Continuously assess password policies to ensure they align with security requirements. As technological advancements occur, it may be necessary to adjust password length and complexity mandates to mitigate new attack vectors.
 \item Password Audits: Regularly audit employee or user passwords to ensure compliance with organizational password policies. This process aids in identifying weak passwords that do not conform to the established requirements and mandates users to change them accordingly.
 \item Ongoing Training: Provide employees with consistent training on password security to enhance their awareness regarding security best practices. Topics may include techniques for creating secure passwords and recognizing phishing attacks.
 \item Monitoring New Security Threats: Remain vigilant regarding new developments and trends within the network security domain to adjust password strategies in response to emerging threats. For instance, advancements in quantum computing may pose risks to traditional encryption methods, necessitating the exploration of new solutions.
 \item Password Expiration Policies: Establish a password expiration timeline that requires users to change their passwords after a specified duration. This practice aids in preventing unauthorized access through the exploitation of outdated and potentially weak passwords.
\end{enumerate}

\section{CONCLUSION}
\label{sec::conclusion}

This study thoroughly analyzes public password databases to identify common patterns and characteristics, providing insights for improving password security policies and user practices. It develops a machine learning-based classifier to distinguish between strong and weak passwords by extracting key features such as length, digit count, uppercase/lowercase letters, and special characters. Six algorithms—SVM, logistic regression, neural networks, decision trees, random forests, and stacked models—are evaluated through hyperparameter tuning and validation metrics (accuracy, recall, F1-score). Results demonstrate that decision trees and stacked models outperform others, achieving the highest scores across all metrics, making them optimal for password strength classification. While neural networks and SVMs show competitive performance in some metrics, they lag slightly in the F1 score. Logistic regression and random forests exhibit notably weaker results, suggesting limited suitability for this task. The findings highlight the efficacy of ensemble methods (stacked models) and interpretable models (decision trees) in password security applications, offering a robust tool for enhancing user-generated password strength.


\bibliographystyle{ACM-Reference-Format}
\bibliography{acmart}

\end{document}